# Interference Management via Space and Frequency Domain Resource Partitioning

Yinan Qi[1], David M. Gutierrez-Estevez [1], Mehrdad Shariat[1], Milos Tesanovic[1] and Maziar Nekovee[2]
[1]Samsung Electronics R&D Institute UK, Staines, Middlesex, UK, [2]Sussex University, Brighton, UK
[1]{ yinan.qi, d.estevez, m.shariat, m.tesanovic}@samsung.com, [2]m.nekovee@sussex.ac.uk

*Abstract*— One key requirement for the fifth-Generation (5G) mobile system is the enhancement of cell edge user performance to ensure that every user is supported with consistent experience anywhere in the network. An active interference design to improve anywhere performance, particularly in the low SINR regime, can be achieved by applying a recently proposed new type of modulation scheme, frequency quadrature-amplitude modulation (FQAM), which could change the distribution of interference and therefore improve channel capacity. In this paper, a resource partitioning scheme to support FQAM in interference intensive scenarios is proposed. The proposed scheme slices radio resources into orthogonal parts for QAM and FQAM, respectively, along two different resource dimensions, namely, space and frequency. This can be achieved by incorporating advanced beamforming algorithms, e.g., full-dimension (FD-MIMO), and performing a frequency-based split of FQAM resources to effectively improve the data rate of the edge users experiencing heavy interference. Two resource partitioning mechanisms are proposed followed by complexity analysis. Evaluation results on both space and frequency-based FQAM interference management are presented, showing significant performance improvements when compared to the regular QAM modulation.

## I. Introduction

One of the pivotal objectives of 5G is to provide a high user experienced rate everywhere, even for the cell edge users experiencing high level of interference. With this ambitious goal set for 5G, use cases requiring 50+ or 100+ Mbps everywhere have been defined in the work of many European projects, e.g. METIS-I and II [1]-[2] and other bodies such as ITU-R and NGMN [3]-[4]. To support such service requirements, a highly flexible 5G interference management combined with advanced novel air-interface design is required to answer this demand. In this regard, an enabling air-interface technique – FQAM – has been developed and actively investigated, showing significant performance enhancement for cell edge users.

QAM modulation has been widely employed in many wire/wireless standards, such as LTE, WiFi [5]. It is well known that the inter-cell interference (ICI) in conventional cellular networks employing orthogonal frequency division multiple-access (OFDMA) with QAM tends to approach a Gaussian distribution [6]. In [7], FQAM was proposed for a downlink cellular OFDMA network to replace conventional QAM and it has demonstrated a significant performance gain for the interference intensive scenario, where the gain is attributed to the fact that the distribution of the ICI plus additive noise (hereafter denoted as ICI for simplicity) received by the user deviates from Gaussian distribution [7], [8].

However, the distribution of ICI when FQAM is applied highly depends on the number of aggressors as shown in [7]. The fewer the number of aggressors, the larger the deviation from Gaussian distribution is, resulting in a greater performance improvement at the victim cell. With increased number of aggressors, the distribution of FQAM ICI asymptotically approaches Gaussian distribution according to the central limit theorem [9]. Thus, the capacity improvement is no longer significant. In addition, in order to make the ICI deviate from Gaussian distribution as far as possible, all aggressors should employ FQAM. However, as identified in [7], the performance improvement can only be achieved when the user equipment (UE) experiences high level of interference from aggressors, i.e., when the UE is in the low SINR regime, and the aggressors switch to FQAM. For those UEs experiencing medium or low level of interference, QAM modulation outperforms FQAM. In this regard, when there are multiple UEs with different interference levels but co-existing in the same cellular network and sharing the same resources, the resources allocated to FQAM and QAM should be orthogonal.

In this paper, we propose an interference management framework based on FQAM and the partitioning of the resource space into two dimensions, namely space and frequency. In the space dimension, the scheme works by combining FQAM with beamforming algorithms. Beamforming is a signal processing technique used in wireless communications for directional signal transmission or reception. Multiple beams can be formed to transmit multiple data streams orthogonal to each other and then different modulation schemes can be employed separately. Therefore, beams can also be regarded as resources, which creates a new degree of freedom to dimension FQAM and QAM resources in an orthogonal manner. While in the frequency domain, dedicated spectrum is allocated to FQAM transmission in interfering cells where the UEs in victim cells will be served from.

The remainder of the paper is organized as follows. In Section II, a brief review of FQAM is presented. In section III, the main resource partitioning concept is overviewed and particularized for space and frequency dimensions. Section IV proposes resource partitioning algorithms in the space and frequency domains. Numerical results are presented in section V. Finally, section VI concludes the paper.

## II. BRIEF FQAM OVERVIEW

An M-ary FQAM modulation scheme is considered formed by combining $M_F$-ary FSK modulation with $M_Q$-ary QAM modulation. One active tone among the $M_F$ tones is selected and modulated with an $M_Q$-ary QAM constellation, yielding a modulation order of $M = M_F M_Q$ also referred to as $(M_F, M_Q)$-FQAM. Each FQAM symbol carries $N = \log_2 M$ information bits. The frequency tones allocated to each of the FQAM symbols are denoted by $f_m$. Further, if we denote the $M_Q$ QAM symbols by $\{x_1, x_2, ..., x_{M_Q}\}$ where $x_k = A_k e^{j\phi_k}$, then the complex baseband equivalent of the FQAM signal at the transmitter can be expressed as follows:

$$x_{m',k}^{FQAM}(t) = \sqrt{2\bar{S}} A_k \delta_{m,m'} \exp(j(2\pi f_m t + \phi_k)) p_T(t) \quad (1)$$

where $m'$ and $m = 0, 1, ..., M_F - 1$, $k = 1, 2, ..., M_Q$, $\bar{S}$ is the average transmit signal power, $p_T(t) = 1$ for $t \in [0, T)$ and zero otherwise, and $\delta_{m,m'}$ is the Kronecker delta function defined as

$$\delta_{m,m'} = \begin{cases} 1, & m = m' \\ 0, & m \neq m' \end{cases} \quad (2)$$

As previously mentioned, when FQAM is applied at interfering cells the interference distribution is shown to deviate from a Gaussian distribution. For a derivation of the ICI-Plus-Noise PDF as well as the soft-decoding decision metrics the reader is encouraged to refer to [7], where an in-depth analysis of FQAM and its performance is presented.

## III. RESOURCE PARTITIONING

The concept of resource partitioning can be illustrated in a multiple dimensional resource space as shown in Fig. 1, where the cube represents available resource dimensions. Orthogonal resources should be allocated to FQAM and QAM, respectively, to optimize the overall performance of the network. This is equal to partitioning the resource cube along different dimension, namely space, frequency, or time in the case of this illustration.

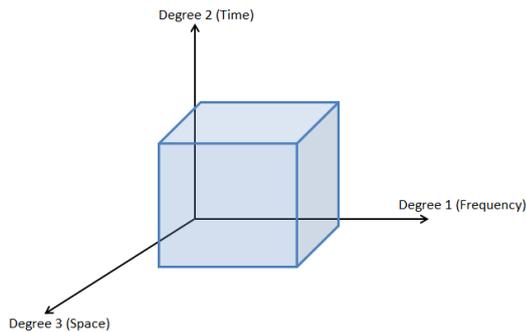

Fig. 1. 3D resource space

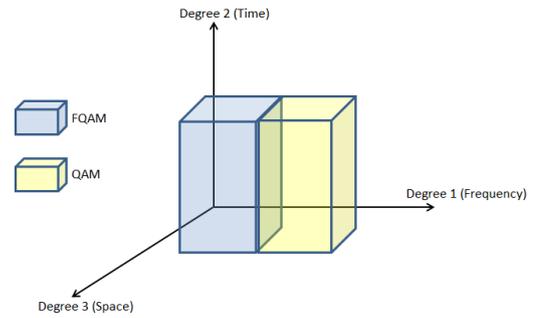

Fig. 2. Frequency dimension partitioning

A sample resource partitioning is depicted in Fig. 2, this case in the frequency domain where dedicated spectrum is allocated to FQAM. To achieve the benefits of FQAM in the cell edge of victim cells while maintaining high throughput in interfering cells, the network should schedule low-SINR users from a flexible and adaptive reserved FQAM frequency resource pool, previously negotiated between neighboring cells. However, this is more applicable in omni-directional transmission where this 3-D space can be simplified into a 2-D space with time and frequency dimensions only. Another option is resource partitioning in the space domain where dedicated beams are allocated to QAM and FQAM, respectively. Nevertheless, all beams occupy the same spectrum simultaneously, which can be realized by multi-stream MIMO, e.g., FD-MIMO. We will present numerical analysis for both space and frequency domains. Furthermore, partitioning in the time domain could also be possible. Moreover, it is worth mentioning that the dimensionality of this resource space can be even larger, and therefore the number of degrees of freedom can be further extended from the illustrated three dimensions to new dimension such as code domain. However, in this paper, we mainly focus on space and frequency domain partitioning.

### A. Partitioning in Space Domain

In the space domain, the proposed resource partitioning concept can be easily applied in the following scenarios. Firstly, all beams from all cells causing interference to their respective neighbouring cells would employ FQAM to contribute to a deviation of the interference distribution from a Gaussian distribution. However, it may not, necessarily, optimize the performance of the entire system since even though the cell edge users may benefit from the employment of FQAM, other users located near the base stations (BSs) should still apply QAM modulation to achieve higher throughput. In this regard, FQAM should be applied to only a subset of the interfering beams in the network with the purpose of maximizing spectral efficiency of the entire system, where the number and selection of beams utilizing FQAM could be obtained using many different methods, such as:

- From interference nulling perspective, FQAM can be applied to beams whose 'victim' cannot null the interference. One example of this is the case in which there are a certain number of affected ('victim') users in the 'victim' cells who cannot use beamforming but have omnidirectional receivers instead. Another example is the case where the 'victim' users could null the interference but are using spatial degrees of freedom for spatial multiplexing;

- From service perspective, FQAM can be applied to beams whose 'victim' is currently using a critical service (e.g. ultra reliable communications).

In practice, space domain partitioning can be employed with advanced beamforming techniques such as FD-MIMO, where orthogonal 2D beams are formed and separate data streams can be transmitted [11].

*B. Partitioning in Frequency Domain*

Interference management via FQAM can be achieved by allocating a dedicated spectrum sub-band to FQAM transmissions in interfering cells where the UEs in victim cells will be served from. An efficient and agile resource management strategy on interference management can then be applied to enable interference control between clusters of mutually interfering cells (or users therein). To achieve the benefits of FQAM in the cell edge of victim cells while maintaining high throughput in interfering cells, low-SINR users are scheduled from a flexible and adaptive reserved resource pool, negotiated between neighboring cells. The details of our proposed interference management scheme are presented in Section IV.B.

*C. Standardization Impact*

New radio (NR) is being standardized in 3GPP and the first version will be frozen by the end of 2017. From RAN1 perspective, one of the main features to be considered is beamforming and areas pertinent to beamforming, such as beam coordination, could potentially enhance the performance of NR significantly. FD-MIMO targets the system utilizing multiple antenna ports at the transmitter side. As aforementioned, advanced beamforming technique opens the new dimension for resource allocation and partitioning. The integration of FD-MIMO and resource partitioning using new modulation schemes such as FQAM could potentially significantly improve the cell edge user performance and therefore the analysis carried out in this paper may have valid impact on 3GPP standardization.

## IV. FQAM PARTITIONING-BASED TECHNIQUES FOR INTERFERENCE MANAGEMENT

In this section, we provide the interference management techniques on the space and frequency resource partitioning dimensions. The model, analysis and algorithms could be easily extended to partitioning in other dimensions.

*A. Interference Management Techniques Using Spatial Partitioning*

As a result of highly directive transmission and reception in beamforming, it is unlikely that a user receives ICI from a large number of interfering small cells, which makes FQAM a suitable solution to tackle the ICI in such cases. When a small cell is able to transmit multiple data streams via multiple beams, different modulation schemes can be used independently depending on the interference situation. E.g. QAM and FQAM are used for high and low interference cases, respectively, as shown in Fig. 3, where each cell transmits using two independent beams to two different UEs, e.g., $S_1$ to $U_{1,1}$ and $U_{2,1}$, respectively. Some of the beams, e.g., the yellow beams, do not cause interferences. On the contrary, the blue beams cause interferences to other UEs when transmitting to the associated UE. Actually, since UEs associated with blue beams are located in the overlapping area, each cell is an aggressor as well as a victim in such a scenario. Thus FQAM can be activated for all blue beams to achieve improved performance.

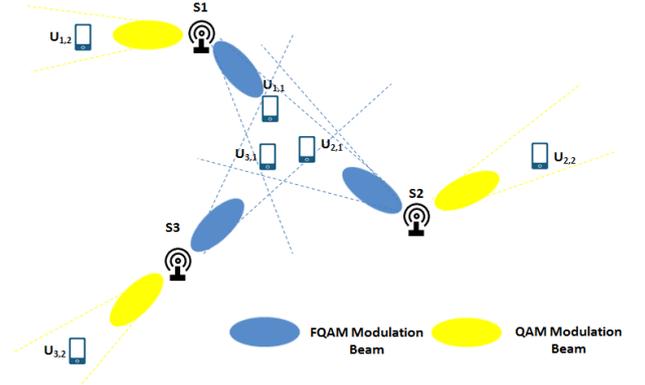

Fig. 3. Multi-beam/user transmission

Fig. 3 represents an interference-symmetric case in the sense that each aggressor is also a victim so that by activating FQAM each BS can benefit from it. However, interference is not always symmetric. The activation of FQAM reduces spectral efficiency in the cell where it is applied in terms of bits/Hz due to the integration of FSK [12]. In the directional transmission scenario, when the interfering beams switch from QAM to FQAM to enhance the rate of the victim UE, the transmission rates to their own associated UEs might be reduced. This is an interference-asymmetric case where an aggressor is not necessarily also a victim at the same time as shown in Fig. 4. As it can be seen, $S_1$ and $S_3$ generate interference to the UE associated with $S_2$ but $S_2$ does not generate interference to the UEs associated with $S_1$ or $S_3$. In order to improve the performance of $U_{2,1}$, $S_1$ and $S_3$ should switch from QAM to FQAM. However, the transmission rates from $S_1$ to $U_{1,1}$ and $S_2$ to $U_{3,1}$ are affected by this switching. In this interference-asymmetric case, the trade-off needs to be taken into consideration to optimize the overall performance.

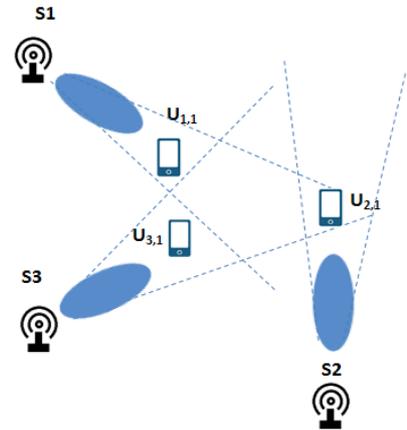

Fig. 4. Interference-asymmetric case

Regarding the procedure itself, in order to switch from QAM to FQAM, a switching threshold should be defined considering two major factors.

- Overall interference level: FQAM should only be activated in interference-intensive scenarios since QAM achieves higher throughput than FQAM in high SINR regime;
- Number of aggressors: as described above, performance improvement of FQAM is due to the derivation from

Gaussian distributed ICI to non-Gaussian distributed ICI. A large number of aggressors will only lead to minor deviation so that the performance improvement is only marginal.

Based on above observations, we define two thresholds: 1) SINR threshold $\gamma_{th}$; and 2) the number of aggressor threshold $N_{th}$. Once the current SINR is below the threshold $\gamma < \gamma_{th}$ and the current number of aggressors is below threshold $N < N_{th}$, QAM can be switched to FQAM. It should be noted that collecting information such as current SINR, number of aggressors, IDs of the aggressors or beams of aggressors requires a dedicated measurement and signalling procedure. The current LTE cell selection/reselection procedure could be useful to identify the overall interference level and a number of main aggressors by measuring signal strength of aggressor reference signals [13]. However, more sophisticated measurement schemes should be developed to facilitate the proposed resource partitioning algorithm but it is out of the scope of this work.

With these two thresholds, we now define the switching mechanism. Before we describe the detailed mechanism, we define two parameters: local service priority level (LSPL) and rate margin (RM). LSPL indicates the priority level of the local service at the cell, which could be a function of the user quality of experience (QoE). For example, for some critical data, LSPL could be very high but for services like web browsing LSPL could be low. This parameter is defined to guarantee that critical services will not be easily interrupted by FQAM switching. In addition, RM indicates how much rate loss a cell can tolerate without sacrificing QoE.

In this paper, the switching mechanism is implemented via a centralized approach that employs a central scheduler coordinating all the BSs by sharing a limited set of information. This makes the algorithm applicable to also non-ideal backhaul between BSs and the central scheduler. Once the UE detects higher interference and low number of aggressors, it informs the victim BS. The victim BS then passes the available information regarding the aggressor/aggressor beams and victim cell/beams to the central scheduler. The central scheduler then makes the final decision with an objective to optimize the global performance of the entire network subject to certain service priorities. Finally, the central scheduler should inform all involved BSs whether or not to activate FQAM for their beams. In a greedy case, the central scheduler could maintain a table of all beams and chooses FQAM/QAM for each beam to optimize the global performance. A similar scheme can be implemented in a distributed manner with local decisions, but due to space constraints we leave its description out of this paper.

In centralized mechanism, optimization is clearly a NP-hard combinational problem that requires very high calculation complexity. However, it can be addressed by some sub-optimal optimization algorithms with much less complexity [14]-[15]. Generally speaking, the network can be managed in a hybrid manner, i.e., it consists of both distributed and centralized mechanisms. The switching between two mechanisms could happen based on current complexity requirements and quality of the backhaul links to reach the balance between performance and complexity.

### B. Interference Management Techniques Using Frequency Partitioning

Our algorithm presented below implements an efficient and flexible resource management strategy on top of FQAM to enable fast yet flexible overhead interference control between clusters of mutually interfering cells (or users therein). The steps of this algorithm are as outlined below:

1. Per target cell, users are split into high-SINR versus low-SINR ones; this can be done e.g. per TTI or on longer intervals.
2. Low-SINR users in a target cell need an active interference management from interfering cells to improve the performance.
3. The interference management can be realized by utilising FQAM in neighbouring interfering cells on resources (in frequency domain) that are reserved for low-SINR users in a target cell.
4. The way of applying different modulation schemes to different frequency ranges can additionally be affected by the type of synchronisation / co-ordination that exists between cells in question.

For the sake of clarity, Fig. 5 illustrates the concept of frequency partition of FQAM, where users with low SINR (i.e. high interference) are allocated the FQAM resources.

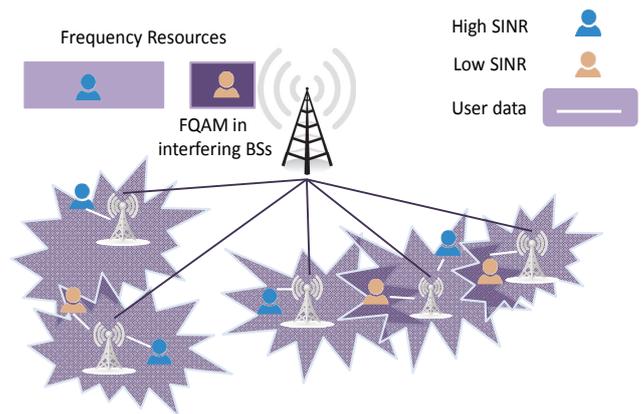

Fig. 5. Interference management based on frequency partition of FQAM

### V. NUMERICAL RESULTS

In this section, we present evaluation results obtained by applying the resource partitioning algorithms in both spatial and frequency domains. The space domain resource partitioning uses beamforming with centralized switching mechanism. Time domain resource partitioning and other dimensions will be evaluated as future work. We assume a synchronous downlink cellular OFDMA network and the long term and short term channel fading models are modelled as in [7].

We assume a homogeneous deployment with 21 BSs as shown in Fig. 6, where each BS is located in the center of one cell. For the spatial partitioning case, each base station uses beamforming to serve the associated UEs. Multiple beams can be formed at each BS and for simplicity, we assume only one UE is served by one beam. For the frequency partitioning case, omnidirectional antennas with 14 dB antenna gain are assumed. In both cases each UE is allocated the same amount of bandwidth and only first tier interference is taken into account. We investigate the following setups:

- For the spatial partitioning case, we assume multiple beams for each BS and each beam only covers one UE. In addition to the UEs associated with blue beams, there are some other UEs covered by red beams that receive low level of interference (denoted as hybrid: QAM+FQAM). A simplified directional antenna model is assumed for both the fixed mmSC and the moving hotspot [16], given as

$$G(\varphi) = \begin{cases} G_0 - 3.01 \times \left(\dfrac{2\varphi}{\varphi_{-3dB}}\right)^2, & 0 \leq \varphi \leq \varphi_{ml}/2 \\ G_{sl}, & \varphi_{ml}/2 \leq \varphi \leq \pi \end{cases}$$

where

$$\phi_{ml} = 2.6\phi_{-3dB},$$
$$G_0 = 10\log\left(\dfrac{1.6162}{\sin(\phi_{-3dB}/2)}\right)^2,$$
$$G_{sl} = -0.4111\ln(\phi_{-3dB}) - 10.579.$$

Here $\varphi$ is an arbitrary angle within the range $[0, \pi]$, $\varphi_{-3dB}$ is the angle of half-power beamwidth, $\varphi_{ml}$ is the main lobe width in units of degrees, and $G_0$ and $G_{sl}$ are the maximum antenna gain and the side lobe gain, respectively.

- For the frequency partitioning case, we assume two users per cell where one is placed on the cell edge (hence belonging to the low SINR user set as described in the previous section) and the other one is randomly located. The random user is always served with QAM. The low SINR user is first served with QAM and then with FQAM to carry out the performance comparison.

The rest of the evaluation parameters are defined in Table-I.

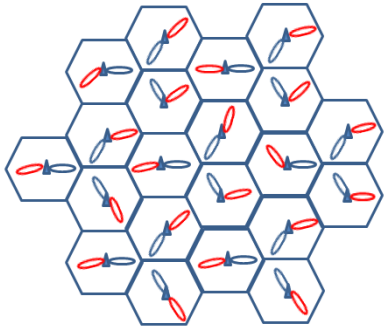

Fig. 6. Deployment for numerical evaluation of FQAM-based interference management (beams are only for spatial partitioning)

Table-I: Simulation Parameters

| Simulation Parameter | Value |
|---|---|
| BS power | 43 dBm |
| Bandwidth allocated to each UE | 20 MHz |
| Number of BSs | 21 |
| Number users per cell | 2 |
| Inter-cell distance | 1732 m |
| Beam width | ($\pi/4$, $2\pi$) |
| Noise temperature | 300 K |

We perform simulations with the target of evaluating the performance of FQAM-based interference management techniques using the space and frequency partitioning techniques described in Section IV. The results show comparisons between an *all-QAM* scenario where all users are served via QAM with a hybrid case where highly interfering users use FQAM and the rest QAM. The metrics that will be evaluated are the 95% available rate (i.e., the lower 5% of the CDF rate curve), the average rate, and the 5% peak transmission rate (i.e., the higher 5% of the CDF). It is worth mentioning that we show results for space and frequency partitioning separately in order not to mislead the reader as their associated all-QAM evaluations correspond to two different scenarios with different results. This is because beamforming introduces gains that omnidirectional antennas cannot provide, and the frequency partitioning case employs a total system bandwidth of 40 MHz as the UE individual bandwidth is 20 MHz and two non-overlapping subbands of that amount of spectrum are employed. A qualitative comparison of the two scenarios is nevertheless provided at the end of this section.

Fig. 7 shows that the 95% available sum-rate using FQAM spatial partitioning can be significantly improved in a hybrid case by applying FQAM to the beams directed to the users experiencing high level of interference. This is because the 95% available rate heavily depends on the users in low SINR regime, i.e., the users associated with blue beams, and FQAM can improve throughput in low SINR regime. Fig. 9 shows the counterpart results for a frequency partitioning case where only the interferers to cell-edge users employ FQAM. A large performance enhancement is also appreciated for the same reason as in the spatial case.

Fig. 9 and Fig. 10 both show the average sum-rate for space and frequency partitionings. As it can be seen, if QAM is applied to all the beams/sub-bands, the average sum-rates are still lower than FQAM. However, applying FQAM does not affect average sum-rate that significantly because mostly only low SINR users experience an increase in throughput while the rest do not get affected. The peak data rates, although not shown, do not experience any difference when the two cases are compared (i.e., QAM and hybrid) since peak rate is normally achieved by those users experiencing low level of interferences and thus QAM should be used.

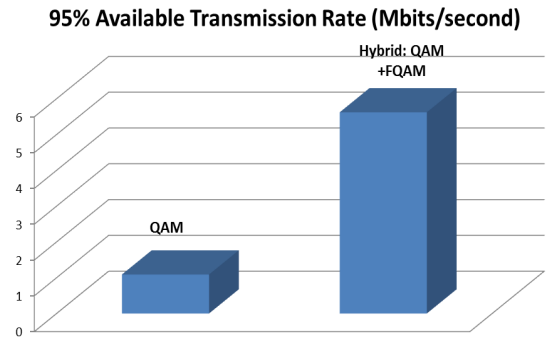

Fig. 7. The 95% available rate with FQAM space partitioning

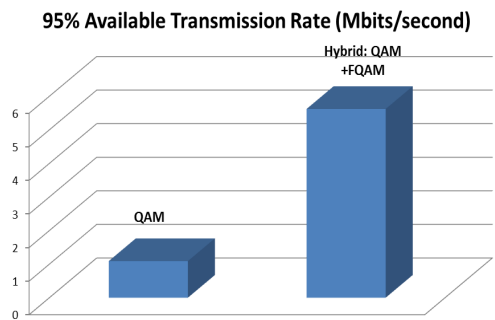

Fig. 8. The 95% available rate with FQAM frequency partitioning

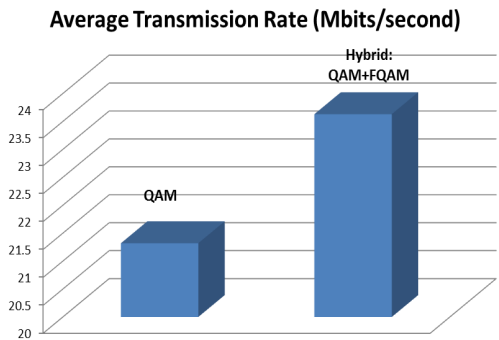

Fig. 9. Average rate with FQAM space partitioning

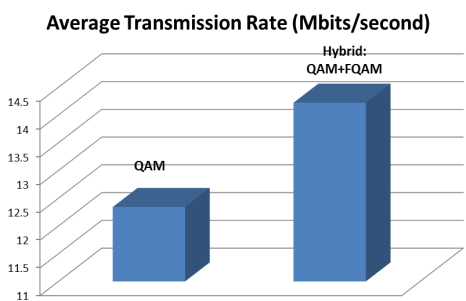

Fig. 10. Average rate with FQAM frequency partitioning

In comparison of space versus frequency FQAM, although the scenarios are not directly comparable, since space partitioning requires additional antennas while frequency partitioning requires a larger system bandwidth. However, the results point to a more spectrally efficient use of the spatial dimension for FQAM, as average transmission rates for spatial partitioning are indeed larger than in frequency partitioning. This is due to the large synergistic benefits of employing beamforming and FQAM simultaneously.

## VI. CONCLUSIONS

In this paper, we identify the orthogonality requirement for the QAM and FQAM resources to maximize the performance of the network. Based on this orthogonality requirement, we propose a novel resource partitioning concept to allocate orthogonal resources to QAM and FQAM for active interference management to enhance the cell edge user performance, ensuring that every user is supported with consistent experience anywhere in the network. The partitioning can be performed in a multi-dimension space, and we focus in the space and frequency dimensions. The space domain partitioning can be effectively combined with advanced beamforming schemes and beam scheduling while frequency partitioning is carried out by assigning a dedicated spectrum subband to FQAM users. Potential for standardization impact is identified, with evaluation results demonstrating significant improvement of the proposed resource partitioning schemes with respect to the conventional QAM modulation, particularly for cell-edge users.

For the future work, we will extend our analysis to time and other dimensions and design efficient measurement signaling procedures to facilitate the proposed resource partitioning schemes.

ACKNOWLEDGEMENTS

This work has been performed in the framework of the H2020 project METIS-II co-funded by the EU. The views expressed are those of the authors and do not necessarily represent the project. The consortium is not liable for any use that may be made of any of the information contained therein.